\documentclass[10pt,twocolumn,letterpaper]{article}

\usepackage{cvpr}
\usepackage{graphicx}
\usepackage{amsmath}
\usepackage{amssymb}
\usepackage{booktabs}
\usepackage{cite}
\usepackage{amsfonts}
\usepackage{bm}
\usepackage{algorithmic}
\usepackage{textcomp}
\usepackage{xcolor}
\usepackage{url}
\usepackage{balance}

\usepackage[pagebackref,breaklinks,colorlinks]{hyperref}

\usepackage[capitalize]{cleveref}
\crefname{section}{Sec.}{Secs.}
\Crefname{section}{Section}{Sections}
\Crefname{table}{Table}{Tables}
\crefname{table}{Tab.}{Tabs.}

\newcommand{\myparatight}[1]{\smallskip\noindent{\bf {#1}:}~}

\def\cvprPaperID{11} 
\def\confName{FedVision}
\def\confYear{2022}

\begin{document}

\title{MPAF: Model Poisoning Attacks to Federated Learning based on Fake Clients}

\author{Xiaoyu Cao\\
Duke University\\
{ xiaoyu.cao@duke.edu}
\and
Neil Zhenqiang Gong\\
Duke University\\
{neil.gong@duke.edu}
}

\date{}
\maketitle

\begin{abstract}
Existing model poisoning attacks to federated learning assume that an attacker has access to a large fraction of compromised genuine clients. However, such assumption is not realistic in production federated learning systems that involve millions of clients. In this work, we propose the first \underline{M}odel \underline{P}oisoning \underline{A}ttack based on \underline{F}ake clients called MPAF. Specifically, we assume the attacker injects  fake clients to a federated learning system and sends carefully crafted fake local model updates to the cloud server during training, such that the learnt global model has low accuracy for many indiscriminate test inputs. Towards this goal, our attack  drags the global model towards an attacker-chosen \emph{base model} that has low accuracy. Specifically, in each round of federated learning,  the fake clients craft fake local model updates that point to the base model and scale them up to amplify their impact before sending them to the cloud server. Our experiments show that MPAF can significantly decrease the test accuracy of the global model, even if classical defenses and norm clipping are adopted, highlighting the need for more advanced defenses. 
\end{abstract}

\section{Introduction}

Federated learning (FL) is an emerging machine learning paradigm for multiple clients (e.g., smartphones or IoT devices) to jointly learn a model with the help of a cloud server. Instead of sharing their private local training data with the cloud server, the clients maintain \emph{local models} to fit their local training data and iteratively share \emph{local model updates} with the cloud server, which aggregates the clients' local model updates to obtain  \emph{global model updates} and uses them to update a \emph{global model}. FL has attracted growing attention in both academia and industry. For instance, Google adopts FL in its Gboard application for next-world prediction \cite{gboard}; a union of world's leading pharmaceutical companies uses FL for drug discovery in a project called MELLODDY \cite{melloddy}; and WeBank leverages FL to predict credit risk of borrowers \cite{webank}. 

However, due to its distributed nature, FL is fundamentally vulnerable to \emph{model poisoning attacks}~\cite{bagdasaryan2020backdoor,bhagoji2019analyzing,fang2019local,xie2020fall}. All existing model poisoning attacks  assume that an attacker has access to compromised genuine clients and rely on their genuine local training data. Specifically, in all or some FL rounds, the compromised genuine clients first compute  local model updates based on their genuine local training data~\cite{fang2019local} or their poisoned versions~\cite{bagdasaryan2020backdoor,bhagoji2019analyzing}, and then further manipulate the local model updates before sending them to the cloud server.  As a result, the learnt global model misclassifies many indiscriminate test inputs  (known as \emph{untargeted  attacks}) \cite{fang2019local} or attacker-chosen ones (known as \emph{targeted  attacks}) \cite{bagdasaryan2020backdoor,bhagoji2019analyzing}. In this work, we focus on untargeted attacks because they are harder to perform as they need to influence the predictions for many indiscriminate test inputs. Existing untargeted model poisoning attacks have shown their effectiveness to FL, even with the presence of Byzantine-robust defenses \cite{Yin18,Blanchard17,Mhamdi18}, i.e., they can reduce the test accuracy of the learnt global model by a significant amount. 

However, existing untargeted model poisoning attacks all require a large fraction of compromised genuine clients and are less effective when the fraction of compromised genuine clients is small~\cite{fang2019local}. 
A recent work \cite{shejwalkarback} argued  that such requirement of a large fraction of compromised genuine clients is not realistic in production FL that involves millions of clients. Specifically,  the cost for compromising genuine clients is so high that an attacker cannot afford to compromise a large fraction of genuine clients in production FL.  For instance, to compromise genuine clients, an attacker needs to pay for the access to a large number of undetected zombie devices. As a result, the fraction of compromised genuine clients is usually small (e.g., 0.01\%) in production FL. Moreover, only a subset of clients are selected in each round of production FL to participate in the training. Therefore, it is likely that no compromised genuine client is selected in many rounds of production FL. Based on these arguments, they came to a conclusion that production FL with the non-robust FedAvg \cite{McMahan17} or classical defenses (e.g., Trimmed-mean \cite{Yin18}) is robust enough against untargeted model poisoning attacks. However, as we will show later, this conclusion does not stand when the attacker can inject fake clients into FL systems and perform model poisoning attacks based on them.

\myparatight{Our work} In this work, we introduce MPAF, the first model poisoning attack to FL that is based on fake clients. We note that the cost of injecting fake clients is much lower than compromising genuine clients in FL. Specifically, the attacker can emulate many fake clients (e.g., android devices) easily using open-source projects \cite{simulatefake} or android emulators \cite{NoxPlayer,bluestacks} on their own machines. 

However, a key challenge of model poisoning attacks based on fake clients is that the fake clients provide no extra knowledge (e.g., no genuine local training data) about the FL system, beyond the global models they receive from the cloud server during training. All existing model poisoning attacks \cite{fang2019local,xie2020fall} rely on the assumption that the attacker has some degree of extra knowledge about the FL system, e.g., the genuine local training data on the compromised genuine clients. In this work, we consider an extreme case for the attacker, where no extra knowledge about the FL system (e.g.,  genuine local training data,  global learning rate, or even the FL method) is available to the attacker, beyond the global models that fake clients receive during training. We note that in FL, the global model is shared with selected clients in each round, including both genuine clients and fake ones. Therefore, our threat model considers the minimum-knowledge scenario for an attacker. 
    
To address the challenge, we propose MPAF, which crafts fake local model updates based on the global models only. Specifically,  in MPAF, an attacker chooses an arbitrary model (called \emph{base model}) that shares the same architecture as the global model and has low test accuracy. For instance, an attacker could randomly initialize a model as the base model. Our intuition is that if we can force the global model to behave like the base model whose test accuracy is low, then the test accuracy of the learnt global model would likely decrease. Therefore, in each round of FL, the fake clients generate the direction of fake local model updates via subtracting the current global model from the base model. The fake clients then scale up the magnitudes of the fake local model updates to enlarge their impact in the global model update. Our evaluations on multiple datasets and multiple FL methods show that MPAF is effective in reducing the test accuracy of the learnt global model even if classical defenses and norm clipping are adopted. For instance,  on Purchase dataset, MPAF decreases the test accuracy of the global model learnt using Trimmed-mean by 32\% when 10\% fake clients are injected. 

Our contribution can be summarized as follows:

\begin{itemize}
    \item We perform the first study on model poisoning attacks to
    FL based on fake clients. 
    \item We propose MPAF, a novel untargeted model poisoning attack that is based on fake clients and requires no extra knowledge about the FL system beyond the received global models during training.
    \item We evaluate MPAF on multiple datasets and multiple FL methods. Our results
    show that MPAF is effective, even if classical defenses and norm clipping are leveraged as a countermeasure.
\end{itemize}

\section{Related Work}
\subsection{Federated Learning (FL)}
Assume there are $n$ clients in FL, each holding some local training data. These clients aim to collaboratively learn a global model with the help of a cloud server. During training, each client maintains a local model based on its local training data and shares its local model updates with the cloud server. Specifically, in the $t$-th round of FL, the cloud server first sends the current global model $\bm{w}^t$ to all or a subset of clients. Then, the clients who receive the global model fine-tune their local models based on the global model using stochastic gradient descent (SGD) and their local training data. 
The clients then send the local model updates to the cloud  server. The cloud server aggregates the local model updates and updates the global model as follows:
\begin{align}
    \bm{w}^{t+1} \gets \bm{w}^t + \eta\bm{g}^t,
    \label{eq:agg}
\end{align}
where $\eta$ is the \emph{global learning rate}, and $\bm{g}^t$ is the \emph{global model update} in the $t$-th round obtained as follows:
\begin{align}
    \bm{g}^t= \mathcal{A}(\bm{g}_1^t, \bm{g}_2^t, \cdots, \bm{g}_n^t).
\end{align}
Here, $\mathcal{A}$ is the aggregation rule the cloud server uses to aggregate the local model updates, which plays an important role in FL. Different FL methods essentially use different aggregation rules. Next, we will discuss three popular aggregation rules, including the non-robust FedAvg \cite{McMahan17}, and two Byzantine-robust ones, i.e.,  Median \cite{Yin18} and Trimmed-mean \cite{Yin18}.

\myparatight{FedAvg} FedAvg \cite{McMahan17} is the most popular aggregation rule in FL. It calculates the average of the local model updates as the global model update. FedAvg achieves the state-of-the-art performance in non-adversarial settings.

\myparatight{Median} Median \cite{Yin18} is a coordinate-wise aggregation rule. The server sorts the values of each parameter in local model updates and finds the median value as the aggregated value for the corresponding parameter in the global model update. 

\myparatight{Trimmed-mean} Trimmed-mean \cite{Yin18} is another coordinate-wise aggregation rule. For each model parameter, instead of using its median value, Trimmed-mean removes the largest and smallest $k$ values from its sorted values, and then computes the average of the remaining values as the corresponding parameter in the global model update. In Trimmed-mean, $k$ achieves a trade-off between the robustness in adversarial settings and test accuracy in non-adversarial settings. In our experiments, we assume a strong defender who knows the number of fake clients, i.e., $k$ equals to the number of fake clients in each round. 

\subsection{Existing Model Poisoning Attacks to FL}
Various attacks  \cite{bagdasaryan2020backdoor,bhagoji2019analyzing,fang2019local,xie2019dba,xie2020fall} have been proposed to poison the global model in FL, all of which rely on compromised genuine clients. Based on the attacker's goal, they can be divided into two categories: untargeted model poisoning attacks \cite{fang2019local,xie2020fall} and  targeted model poisoning attacks \cite{bagdasaryan2020backdoor,bhagoji2019analyzing,xie2019dba}.
Untargeted model poisoning attacks aim to decrease the test accuracy of the global model, while targeted model poisoning attacks aim to force the global model to output attacker-chosen target labels for attacker-chosen target inputs. We focus on untargeted model poisoning attacks in this work.

Existing untargeted model poisoning attacks  \cite{fang2019local,tolpegin2020data,xie2020fall} follow the following two steps in all or multiple rounds of FL. First, the compromised genuine clients compute the genuine local model updates based on their genuine local training data. Then, they perturb their genuine local model updates such that the poisoned global model updates will substantially deviate from the genuine ones.  These attacks require many compromised genuine clients to be effective. 
However, in production FL, it may not be affordable for an attacker to obtain access to a large number of compromised genuine clients \cite{shejwalkarback}. Therefore, we consider a more practical scenario for model poisoning attacks that an attacker injects fake clients to the FL system. Unfortunately, existing attacks are not applicable to such scenario, since they require extra knowledge about the FL system (e.g., genuine local training data), which is not available on the fake clients. We notice that several works \cite{lin2019free,fraboni2021free} studied the free-rider attacks with fake clients, which are orthogonal to model poisoning attacks. 

\subsection{Defenses against Model Poisoning Attacks}
Many defenses \cite{Blanchard17,ChenPOMACS17,Mhamdi18,rajput2019detox,xie2019zeno,Yin18,fang2019local,shen2016auror} have been proposed against model poisoning attacks to FL, which fall into two main categories. The first type of defense \cite{Blanchard17,ChenPOMACS17,Mhamdi18,rajput2019detox,xie2019zeno,Yin18} designs Byzantine-robust aggregation rules. Their idea is to mitigate the impact of statistical outliers among the local model updates. For instance, Trimmed-mean \cite{Yin18} removes the largest and smallest values of each coordinate in the local model updates before taking the average. The other type of defense \cite{cao2021provably, xie2021crfl} aims to provide provable guarantee against poisoning attacks. For instance, Cao et al. \cite{cao2021provably} leveraged the fundamental robustness of majority vote to design an ensemble-based provably secure federated learning framework. They proved that when the number of compromised genuine clients is bounded, the predictions for test inputs are not affected by any attack. However, their derived provable security guarantee does not consider fake clients. 

A recent work \cite{shejwalkarback} claims that production FL with the non-robust FedAvg or classical defenses such as Trimmed-mean is already robust against  untargeted model poisoning attacks that rely on compromised genuine clients, because the fraction of compromised genuine clients is small in production FL systems. 
However, this claim does not stand for fake clients based model poisoning attacks. As we will show, an attacker can inject many fake clients into FL systems and perform MPAF to degrade the performance of the learnt global model. 

In fact, the claim on the robustness of FedAvg is not accurate even if the attacker only has access to  a small fraction of compromised genuine clients. \cite{shejwalkarback} claims that FedAvg is robust because 1) the server selects a small fraction of clients in each global training round, 2) compromised genuine clients are unlikely to be selected when their fraction is small, and 3) the compromised genuine clients' impact on the global model will be eliminated during training even if they are selected in certain training rounds. However, robustness/security is about an FL system's performance in the \emph{worst-case} scenarios.  A compromised genuine client can substantially degrade the global model's accuracy in the scenario where it is selected near the end of the training process. Although such worst-case scenario happens with a small probability when the fraction of compromised genuine clients is small, it still invalidates the robustness of FedAvg.

\section{Threat Model}

\subsection{Attacker's Goal}
The attacker's goal is to decrease the test accuracy of the learnt global model. Specifically, a larger difference between the test accuracy of the global models with and without attack indicates a stronger attack.

\subsection{Attacker's Capability}
We assume the attacker can inject many fake clients into FL systems. The attacker can control these fake clients to send arbitrary fake local model updates to the cloud server.  

Compared to compromising genuine clients, the cost of injecting fake clients is much more affordable. Specifically, to compromise genuine clients, an attacker needs to bypass the anti-malware software on the clients' devices, which becomes more difficult as the anti-malware industry evolves.  The attacker may also choose to pay for the zombie devices that are already compromised and could be remotely accessed. However,  it would be too costly to buy a large number of zombie devices. Moreover, performing the attacks on compromised devices requires the attacker to evade the anomaly detection on the systems, making it even harder. 

On the contrary, it would be easy and cheap to perform attacks based on fake clients. First, an attacker can emulate fake clients using open-source projects \cite{simulatefake}, or even the free softwares, e.g., android emulators on PC \cite{NoxPlayer,bluestacks}. It is worth noting that modern android emulators support multi-instance functionality, which means that an attacker can emulate many instances (clients) using a single machine, significantly reducing the cost. Another advantage of using fake client is that the attacker has full control over the devices. For instance, the android emulators can grant the attacker root access to the devices, and the attacker does not need to deal with any alert that the system may probably raise during the attack. 

\subsection{Attacker's Knowledge}
Existing model poisoning attacks that rely on compromised genuine clients assume the attacker knows extra knowledge about the FL system, e.g., the genuine local training data on the compromised genuine clients, other than the received global models during training. However, such assumption often does not hold when it comes to attacks based on fake clients. Specifically, the fake clients are created by the attacker and there are usually no genuine local training data on them. Therefore, we consider a more realistic threat model, where the attacker has no knowledge about the FL system other than the received global models during training. In particular, the attacker does not know any local training data or local model updates on any genuine client. Moreover, the attacker does not know the FL aggregation rule or the global learning rate that the cloud server uses. Since the global model is broadcast to selected clients in each round of FL, including both genuine and fake clients,  our threat model considers the scenario with minimum knowledge for the attacker. 

\section{Our Attack}
We will first discuss two baseline attacks and analyze why they are not effective. Then, we will introduce our MPAF.

\subsection{Baseline Attacks}
\label{sec:consistency}
A naive way of performing model poisoning attacks with limited knowledge is to use random noise as the fake local model updates. For instance, the fake clients could sample a  Gaussian random noise for each model parameter. They can then enlarge the magnitudes of the random local model updates using a scaling factor $\lambda$. The fake clients send the scaled random noise to the cloud server as the fake local model updates. Formally, the $i$-th fake client sends $\bm{g}_i^t=-\lambda\mathbf{\varepsilon}$ to the cloud server in the $t$-th round, where $\mathbf{\varepsilon}$ is a random vector sampled from the multivariate Gaussian distribution $ \mathcal{N}(\bm{0}, \bm{I})$.
We call such attack \emph{random attack}.

Another intuitive attack based on fake clients is to estimate the benign global model updates using historical information, and then generate fake local model updates that have the opposite direction. Specifically, in the $t$-th round, given the current global model $\bm{w}^t$ and the previous global model $\bm{w}^{t-1}$, we can compute the global model update $\bm{g}^{t-1}$ in the $(t-1)$-th round as $\bm{g}^{t-1}=(\bm{w}^t-\bm{w}^{t-1})/\eta$, where $\eta$ is the global learning rate. Since the global model updates in consecutive rounds do not differ much, especially when the global model is near convergence, we can approximate the benign global model update in the $t$-th round as $\hat{\bm{g}}^t\approx\bm{g}^{t-1}=(\bm{w}^t-\bm{w}^{t-1})/\eta$.  Under our threat model, the global learning rate $\eta$ is unknown to the attacker. However, an attacker does not need to know the exact magnitude of the benign global model update. Instead, the attacker can use a large scaling factor $\lambda$ to scale up the fake local model updates such that their magnitudes are no smaller than the ones from the genuine clients. Formally, a fake client $i$ sends $\bm{g}_i^t=-\lambda(\bm{w}^t-\bm{w}^{t-1})$ to the cloud server in the $t$-th round, where the negative sign means the attacker aims to deviate the global model to the opposite direction. We call such attack \emph{history attack}.

The two baseline attacks are intuitive. However, as we will show in Section \ref{sec:evaluation}, they have limited impact on the accuracy of the learnt global model when classical defenses (e.g., Trimmed-mean) are applied. We suspect that this is because the attacks are not \emph{consistent} in consecutive rounds. Specifically, the attacks may successfully deviate the global model to some direction by a small step in each individual FL round. However, such deviations may have different directions in different rounds, which means the deviations may cancel out in multiple rounds.

\begin{figure}
    \centering
    \includegraphics[width=0.48\textwidth]{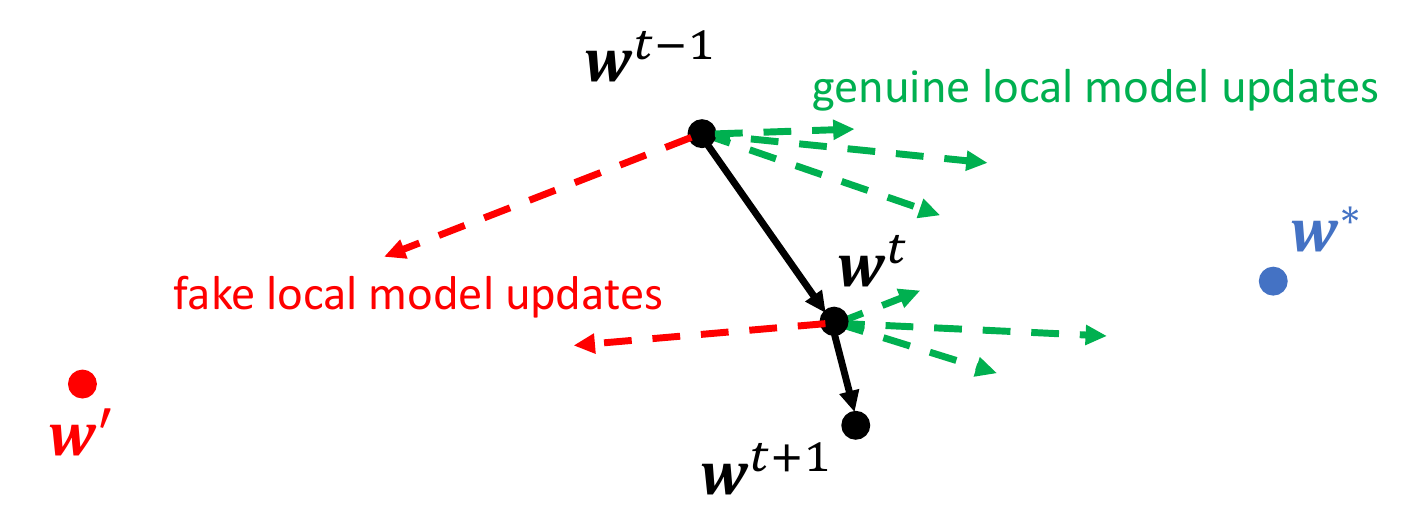}
    \caption{Illustration of MPAF. $\bm{w}'$ is an attacker-chosen base model. $\bm{w}^{t-1}, \bm{w}^t,$ and $\bm{w}^{t+1}$ are the global models in round $t-1, t,$ and $t+1$, respectively.  $\bm{w}^*$ is the learnt global model without attack. The fake local model updates from the fake clients drag the global model towards the base model.}
    \label{fig:illust}
\end{figure}

\subsection{MPAF}
Figure \ref{fig:illust} illustrates our MPAF. The attacker selects a \emph{base model} $\bm{w}'$ that has low test accuracy. For instance,  the attacker can select a randomly initialized model as the {base model}, whose test accuracy is near random guessing. In MPAF, the fake clients craft their local model updates to  drag the global model  towards the base model. Specifically, in the $t$-th round of FL, the fake clients generate fake local model updates, whose direction is determined via subtracting the current global model parameters from the base model parameters. Then the fake clients scale up their fake local model updates by a factor $\lambda$ to amplify their impact.

The key challenge of attacks based on fake clients is that the attacker has minimum knowledge about the FL system, i.e., only the global models received during training.  Therefore, finding an effective way of leveraging such limited information becomes the critical component of attacks. In MPAF, our main idea is to force the global model to mimic the base model $\bm{w}'$. Formally, we formulate our attack as the following optimization problem:
\begin{align}
    \min_{\bm{g}_i^t, i\in[n+1,n+m],t\in[0,T-1]} \Vert\bm{w}^T-\bm{w}'\Vert,
\end{align}
where $n$ is the number of genuine clients, $m$ is the number of fake clients ($n+1, n+2, \cdots, n+m$ are the fake clients),  $T$ is the number of FL rounds during training, $\bm{w}^T$ is the learnt final global model, and $\Vert\cdot\Vert$ represents the $\ell_2$ norm. 
Note that our problem formulation takes the entire training process into consideration. Specifically, in any FL round, the fake clients have the same goal of deviating the final global model towards a fixed attacker-chosen base model.

We solve the optimization problem via driving the global model towards the base model in each FL round. 
Specifically, in the $t$-th round of FL, the fake clients compute the direction of fake local model updates by subtracting the current global model from the base model, i.e., $\bm{d} = \bm{w}'-\bm{w}^t$. The global model is closer to the base model if it is deviated to this direction. Then, the fake clients scale up $\bm{d}$ by a factor $\lambda$ to amplify the magnitude. The final fake local model update for a fake client $i$ in the $t$-th round is as follows:
\begin{align}
    \bm{g}^t_i=\lambda(\bm{w}'-\bm{w}^t).
\end{align}
An attacker can choose a large $\lambda$ to guarantee that the attack is still effective after the cloud server aggregates the fake local model updates from the fake clients and the genuine local model updates from the genuine clients.  

\section{Evaluation}
\label{sec:evaluation}

\subsection{Experimental Setup}
\subsubsection{Datasets and Global Model Architectures}
We evaluate our attacks using multiple datasets, i.e., MNIST \cite{lecun2010mnist}, Fashion-MNIST \cite{xiao2017/online}, and Purchase \cite{purchase}.

\myparatight{MNIST} MNIST \cite{lecun2010mnist} is a benchmark image classification dataset. There are 60,000 training examples and 10,000 testing examples of 10 classes, where each example is a hand-written digit image of size $32\times32$. Following \cite{fang2019local}, we distributed the training examples to the clients with degree of non-IID $q=0.5$ to simulate non-IID training data. We use the same CNN architecture for the global model as in \cite{cao2020fltrust}.

\myparatight{Fashion-MNIST} Like MNIST, Fashion-MNIST \cite{xiao2017/online} is a 10-class image classification dataset with 60,000 training examples and 10,000 testing examples. Similar to MNIST, we distribute the training examples to the clients with degree of non-IID $q=0.5$. We use the same CNN as the one for MNIST.

\myparatight{Purchase} Purchase \cite{purchase} is a 100-class classification dataset, whose goal is to predict customer's purchase styles. There are 197,324 purchase records in Purchase, each of which has 600 binary features. We split the dataset into 180,000 training records and 17,324 test records. We distribute the training data evenly to the clients. We use a fully connected neural network as the global model architecture. There is one hidden layer in the network, whose number of neurons is 1,024 and activation function is Tanh. 

\begin{figure*}[!t]
    \center
    \subfloat{\includegraphics[width=0.33\textwidth]{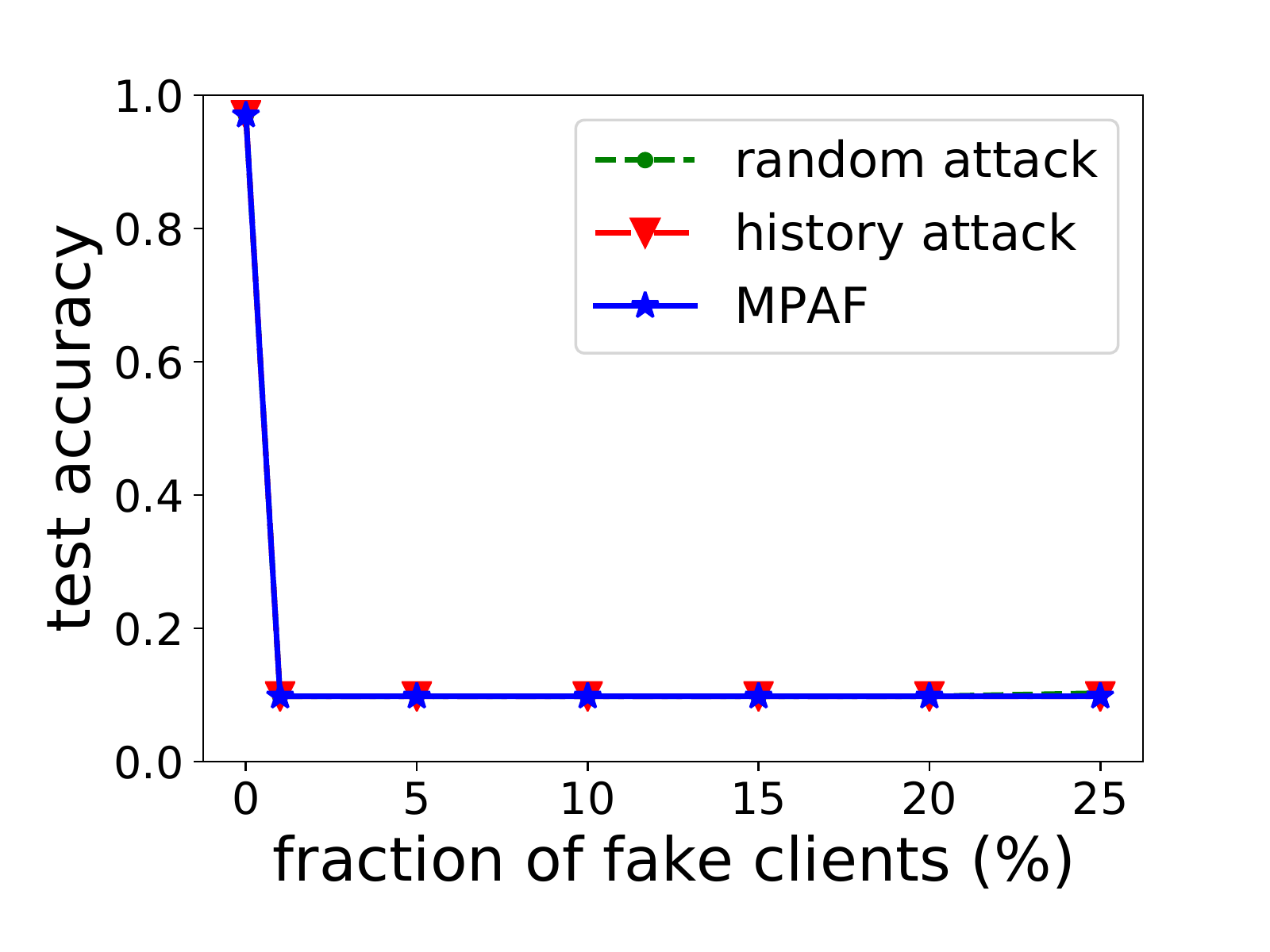}}
    \subfloat{\includegraphics[width=0.33\textwidth]{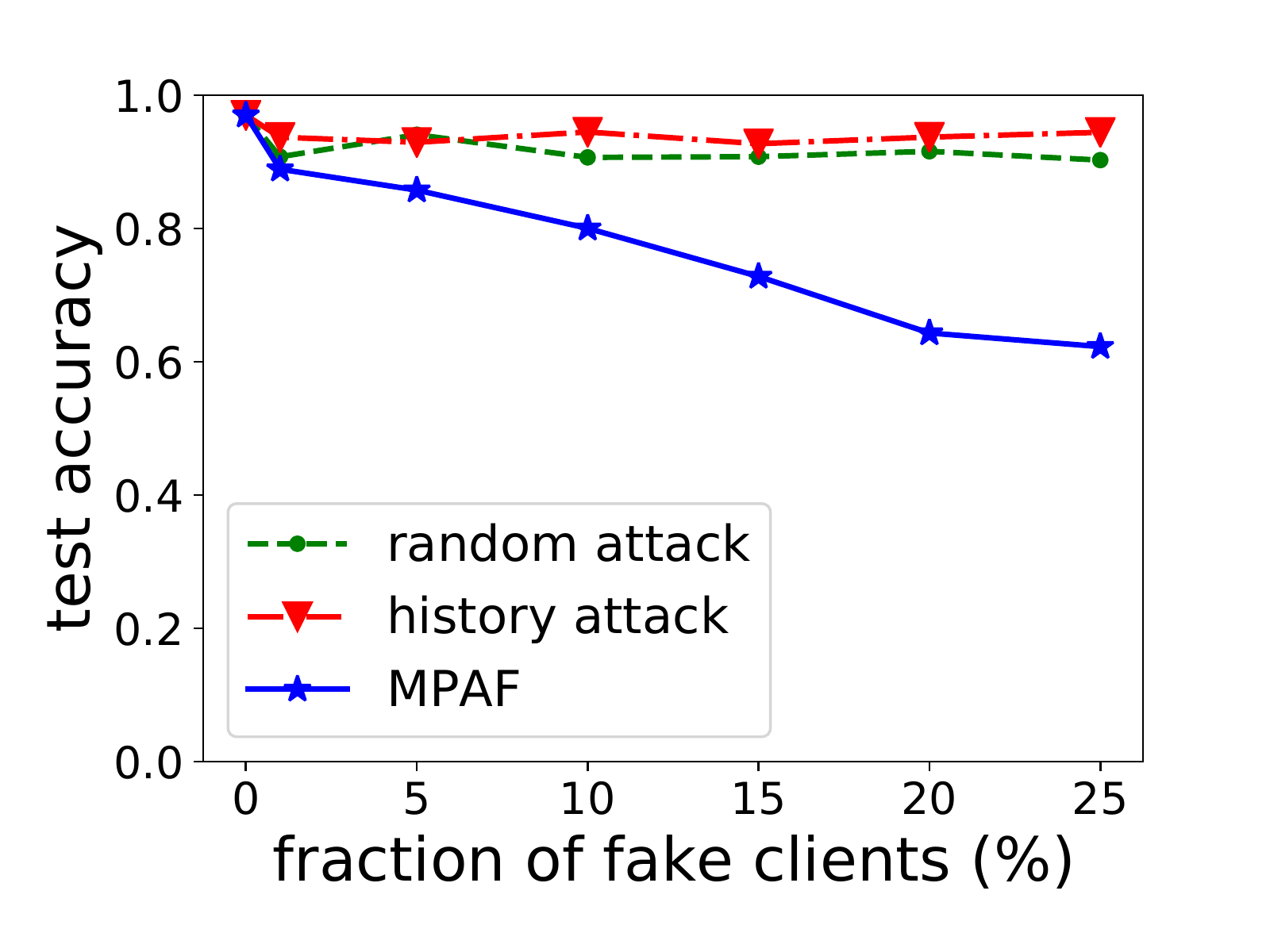}}
    \subfloat{\includegraphics[width=0.33\textwidth]{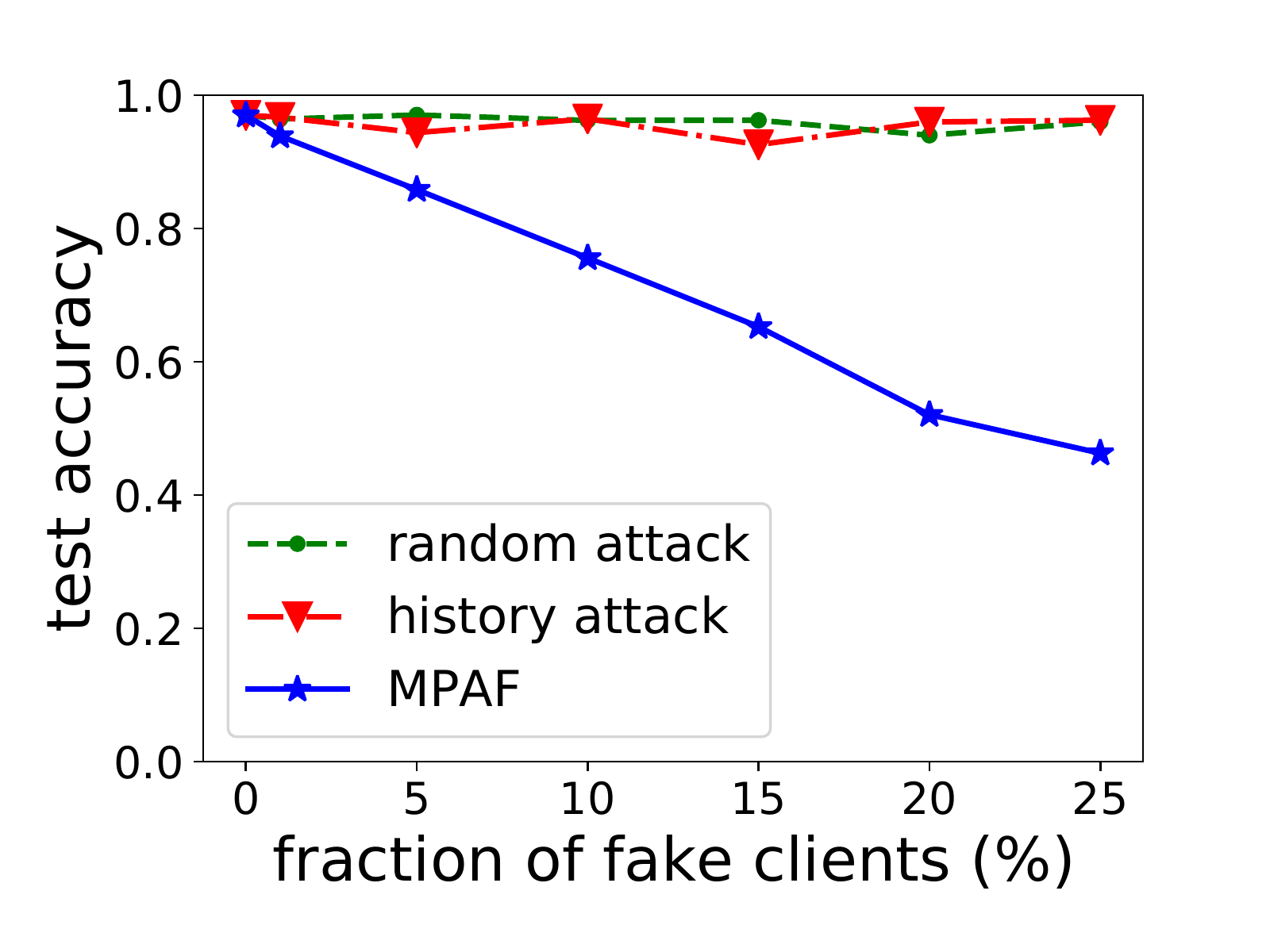}}\\
    \subfloat{\includegraphics[width=0.33\textwidth]{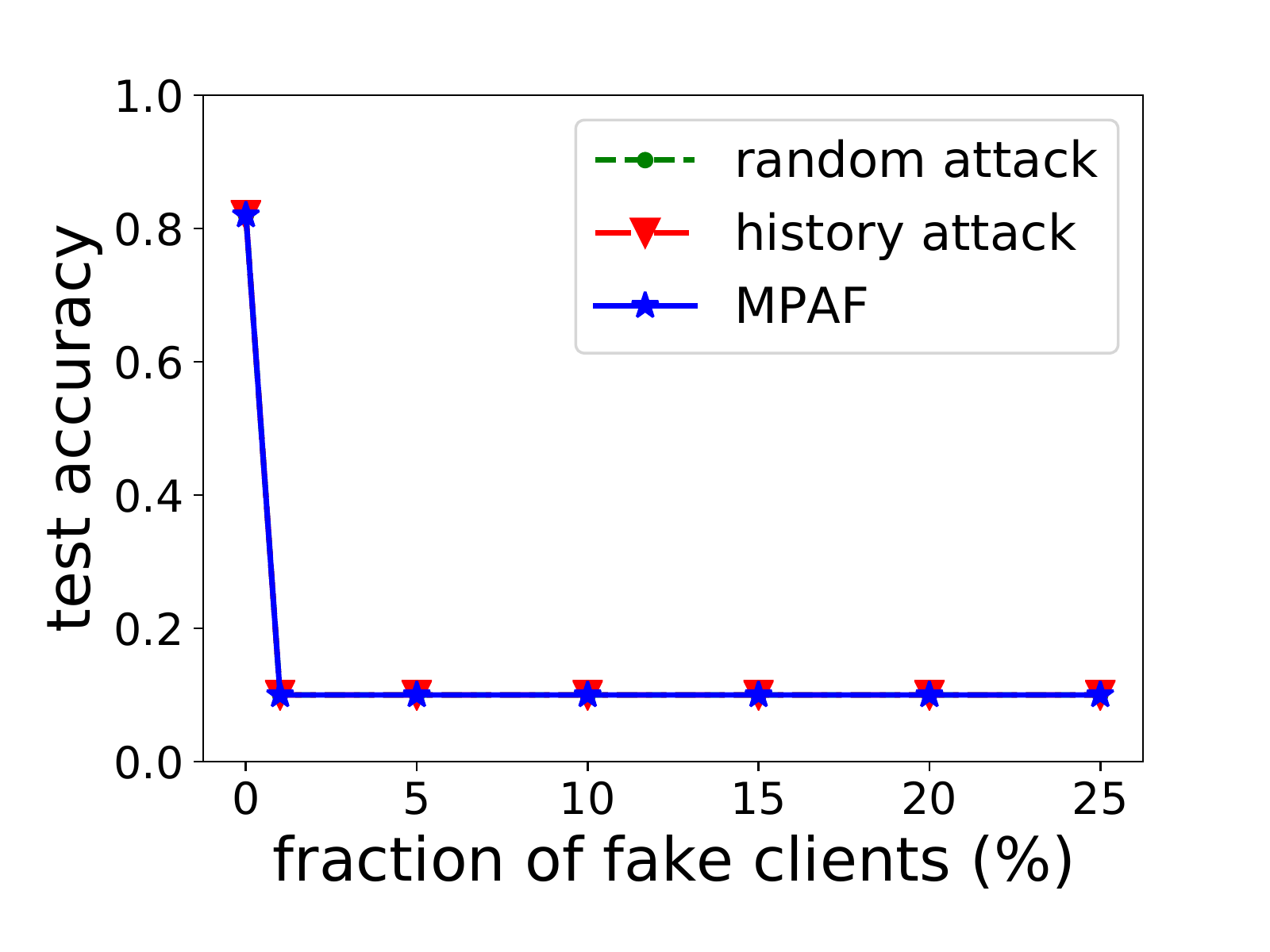}}
    \subfloat{\includegraphics[width=0.33\textwidth]{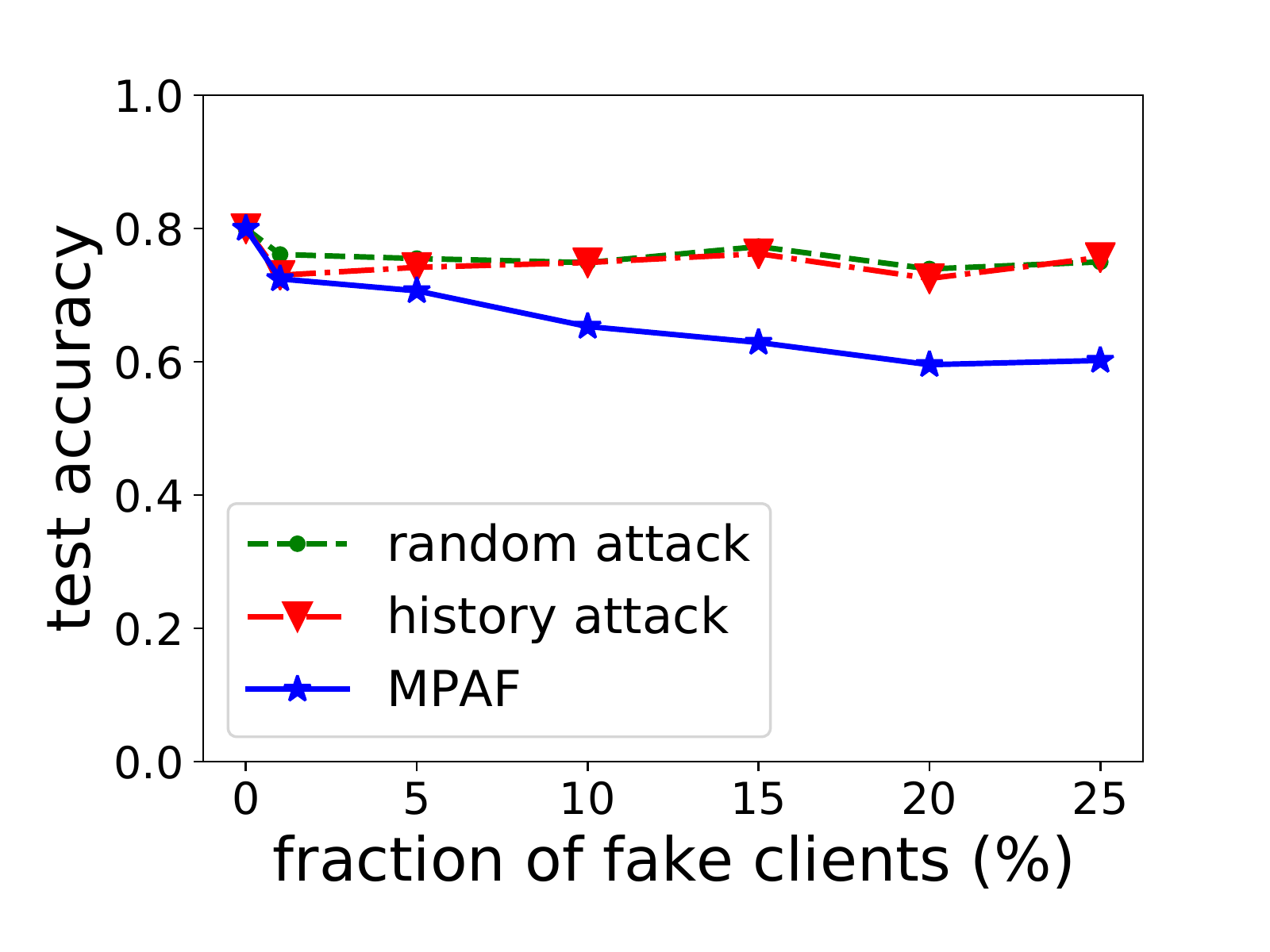}}
    \subfloat{\includegraphics[width=0.33\textwidth]{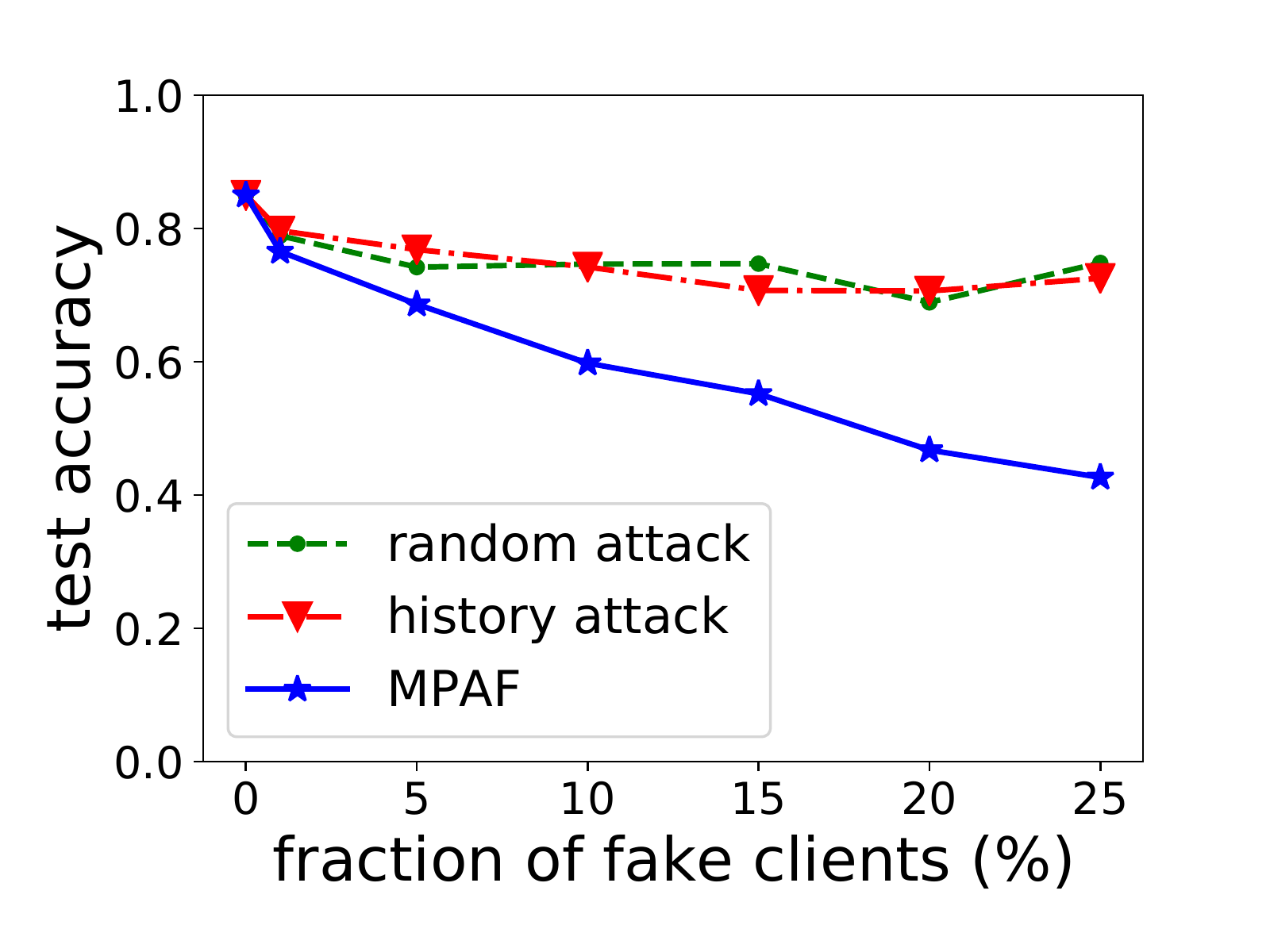}}\\
    \setcounter{subfigure}{0}
    \subfloat[FedAvg]{\includegraphics[width=0.33\textwidth]{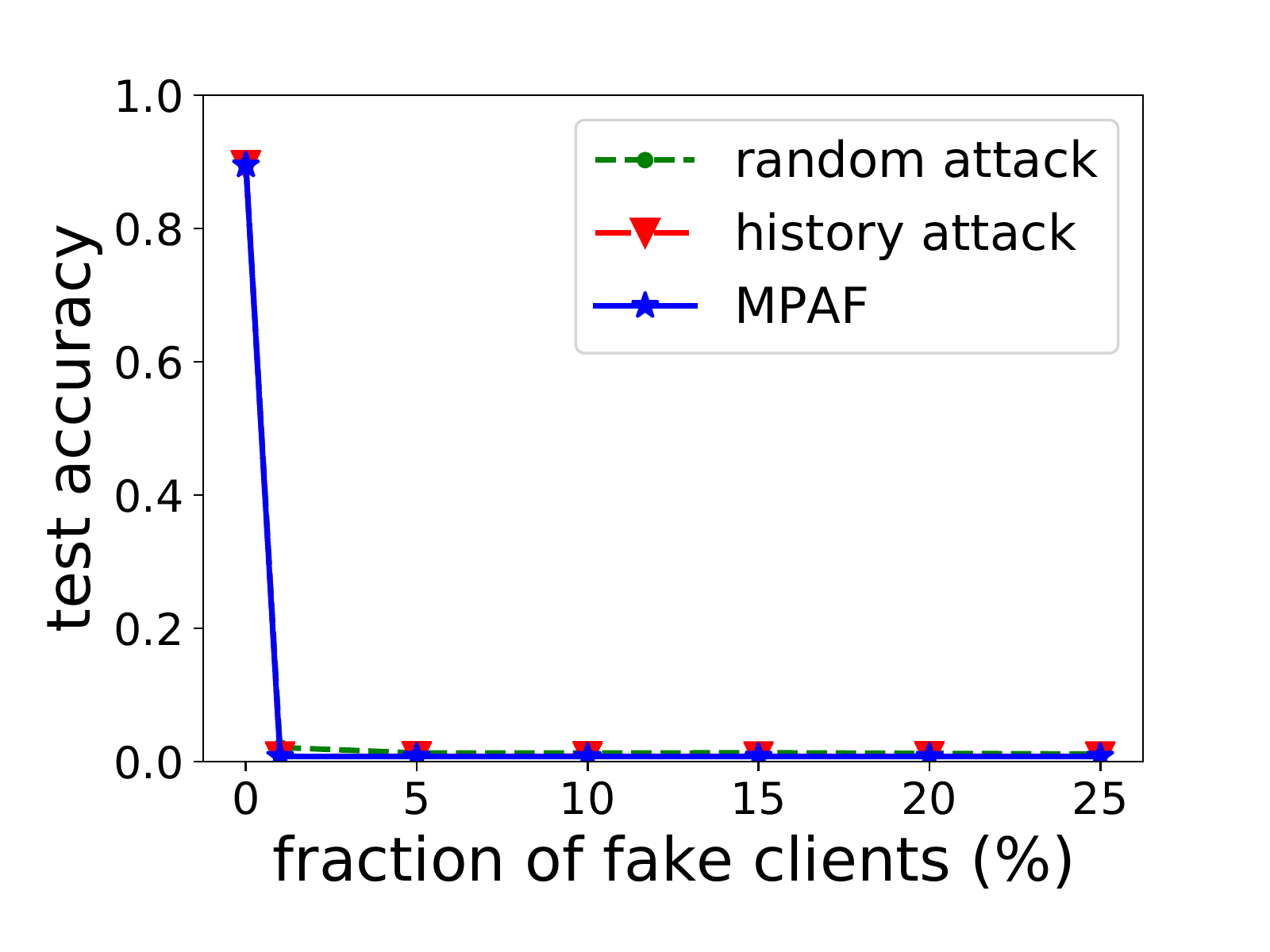}}
    \subfloat[Median]{\includegraphics[width=0.33\textwidth]{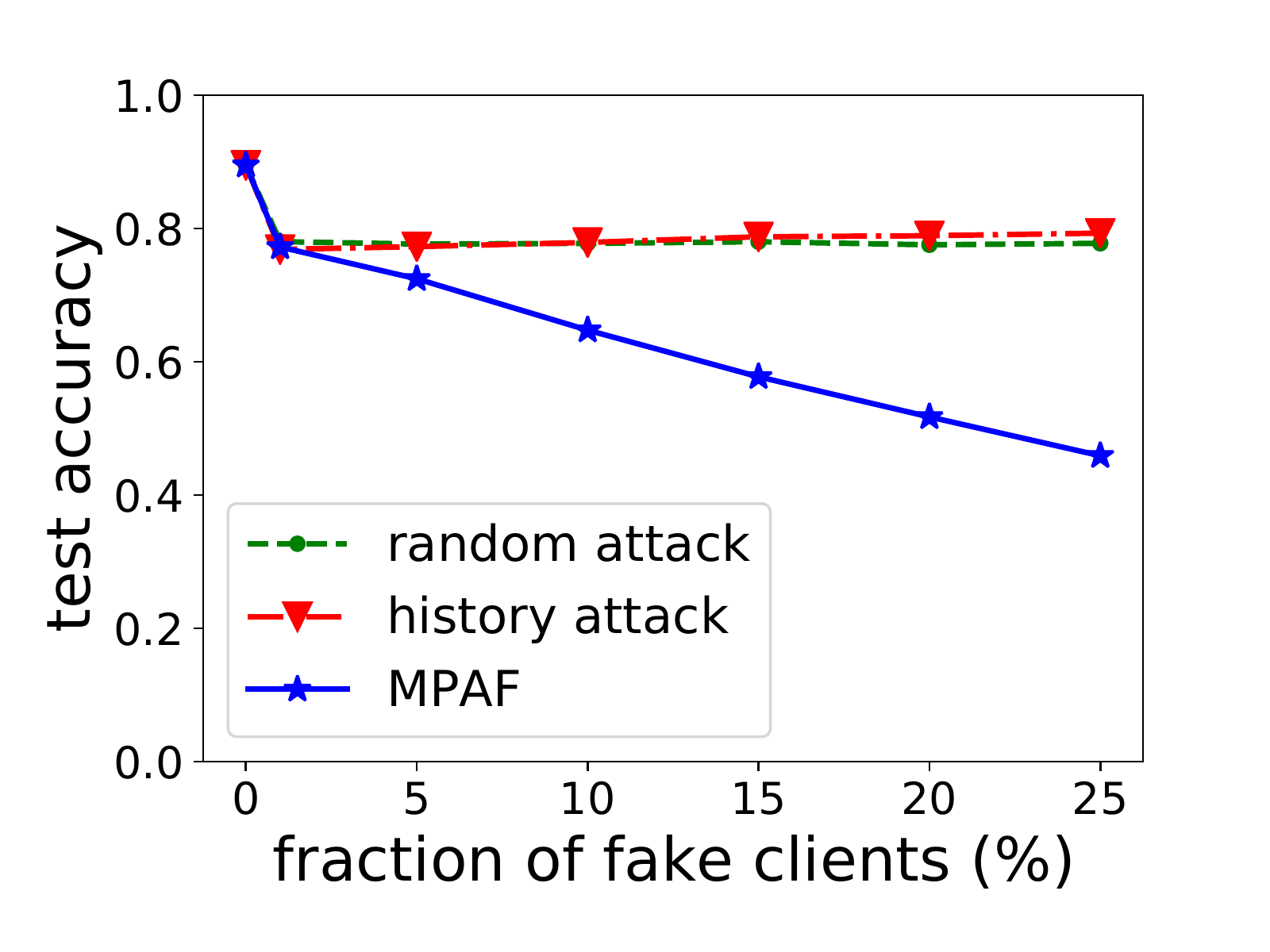}}
    \subfloat[Trimmed-mean]{\includegraphics[width=0.33\textwidth]{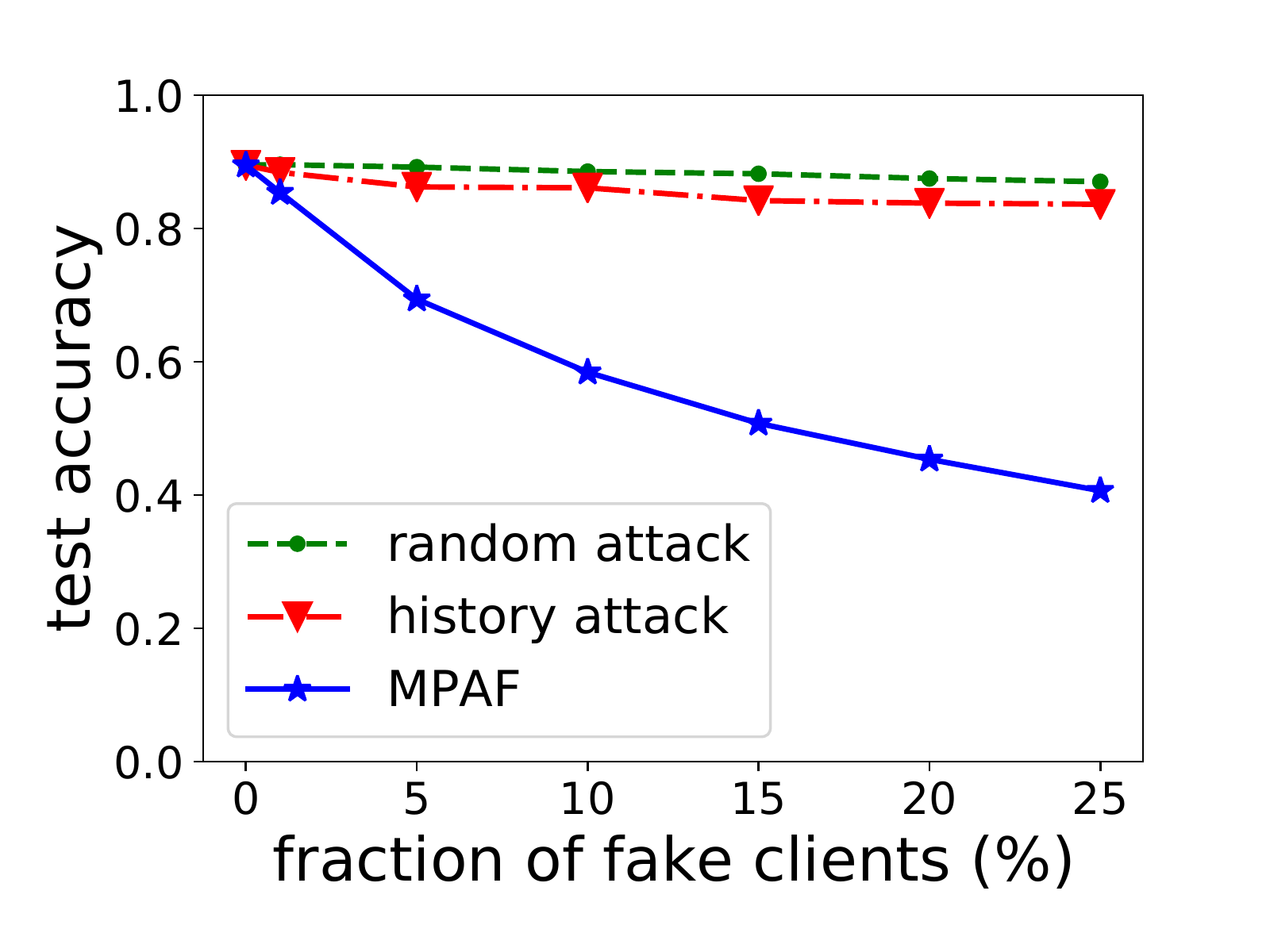}}\\
    \caption{Test accuracy of the global models learnt by different FL methods under the three attacks when the fraction of fake clients varies. The datasets are MNIST (first row), Fashion-MNIST (second row) and
    Purchase (third row).}
    \label{fig:fraction}
\end{figure*}

\subsubsection{FL and Attack Settings}
For all three datasets, we assume there are $n=1,000$ genuine clients in total.  We define the fraction of fake clients as the number of injected fake clients divided by the number of genuine clients, i.e., $m/n$. By default, we assume there are $m=100$ fake clients, i.e., the fraction of injected fake clients is 10\%, unless otherwise mentioned. In each round of FL, the genuine clients train their local model using SGD with batch size of 32, 32, and 128 for MNIST, Fashion-MNIST, and Purchase, respectively. We set the global learning rate $\eta$ to 0.01, 0.01, and 0.005 for the three datasets, respectively. We use different settings for different datasets to achieve high test accuracy in non-adversarial settings.  In each FL round, we assume the cloud server randomly samples $\beta$ fraction of clients to participate in training. We set the default value of $\beta$ to 1, i.e., the server selects all clients in each round during training. We will evaluate the impact of $\beta$ in our experiments. We further set the number of FL rounds $T$ to $\frac{200}{\beta}, \frac{200}{\beta},$ and $\frac{500}{\beta}$ for the three datasets. This is because  a smaller $\beta$ indicates fewer clients in each FL round, thus needs more rounds to converge. For our attacks, we set the default value of the scaling factor $\lambda=1\times 10^6$ and we will explore its impact. We repeat the attacks in each experiment for 20 times with different random seeds and report the average results.

\subsubsection{Evaluation Metric}
We focus on untargeted model poisoning attacks in this work, whose goal is to decrease the test accuracy of the learnt global model. Therefore, we use the test accuracy of the learnt global models as our metric. A lower test accuracy indicates a stronger attack.

\begin{figure*}[!ht]
    \center
    \subfloat[MNIST]{\includegraphics[width=0.33\textwidth]{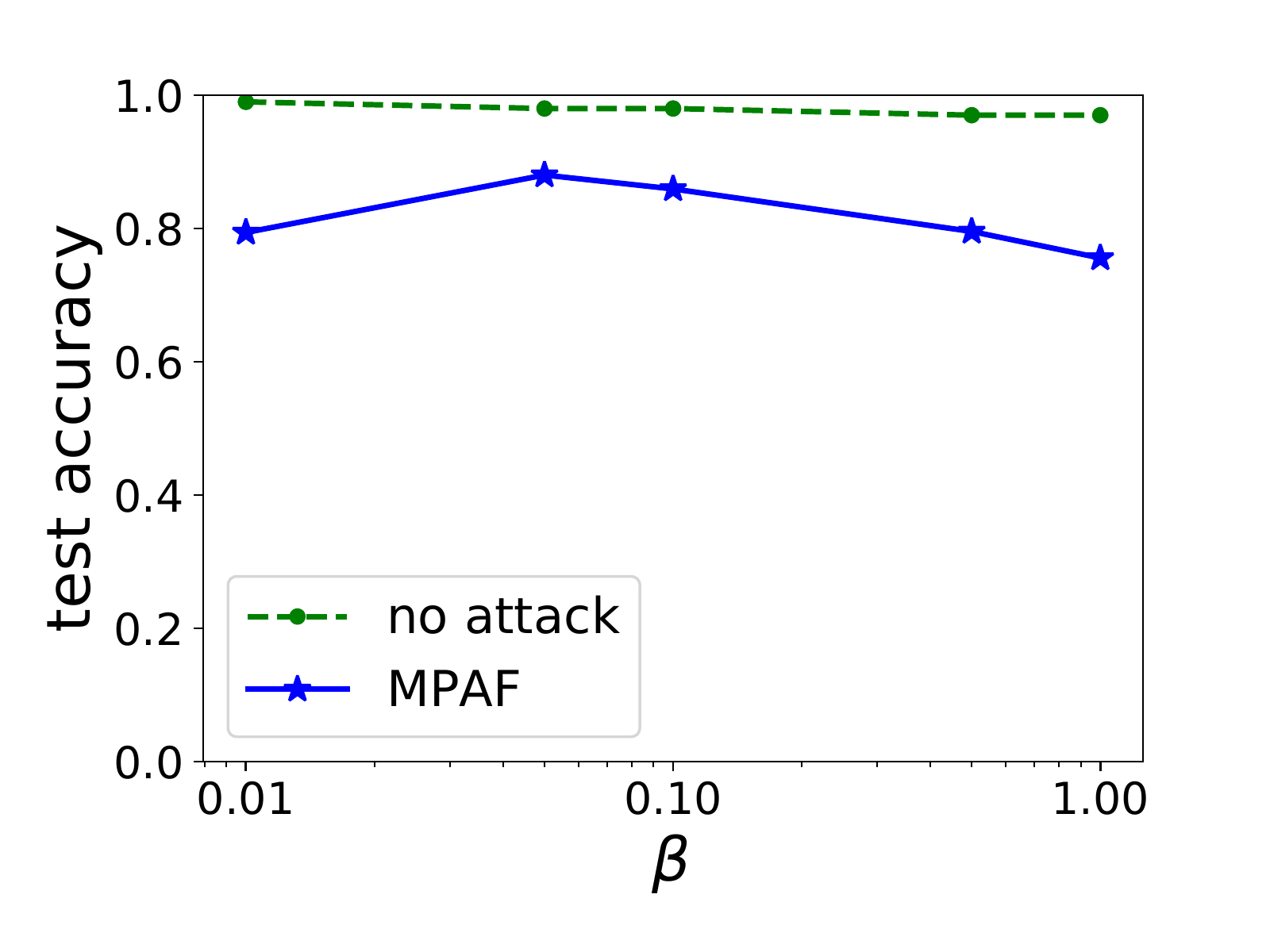}}
    \subfloat[Fashion-MNIST]{\includegraphics[width=0.33\textwidth]{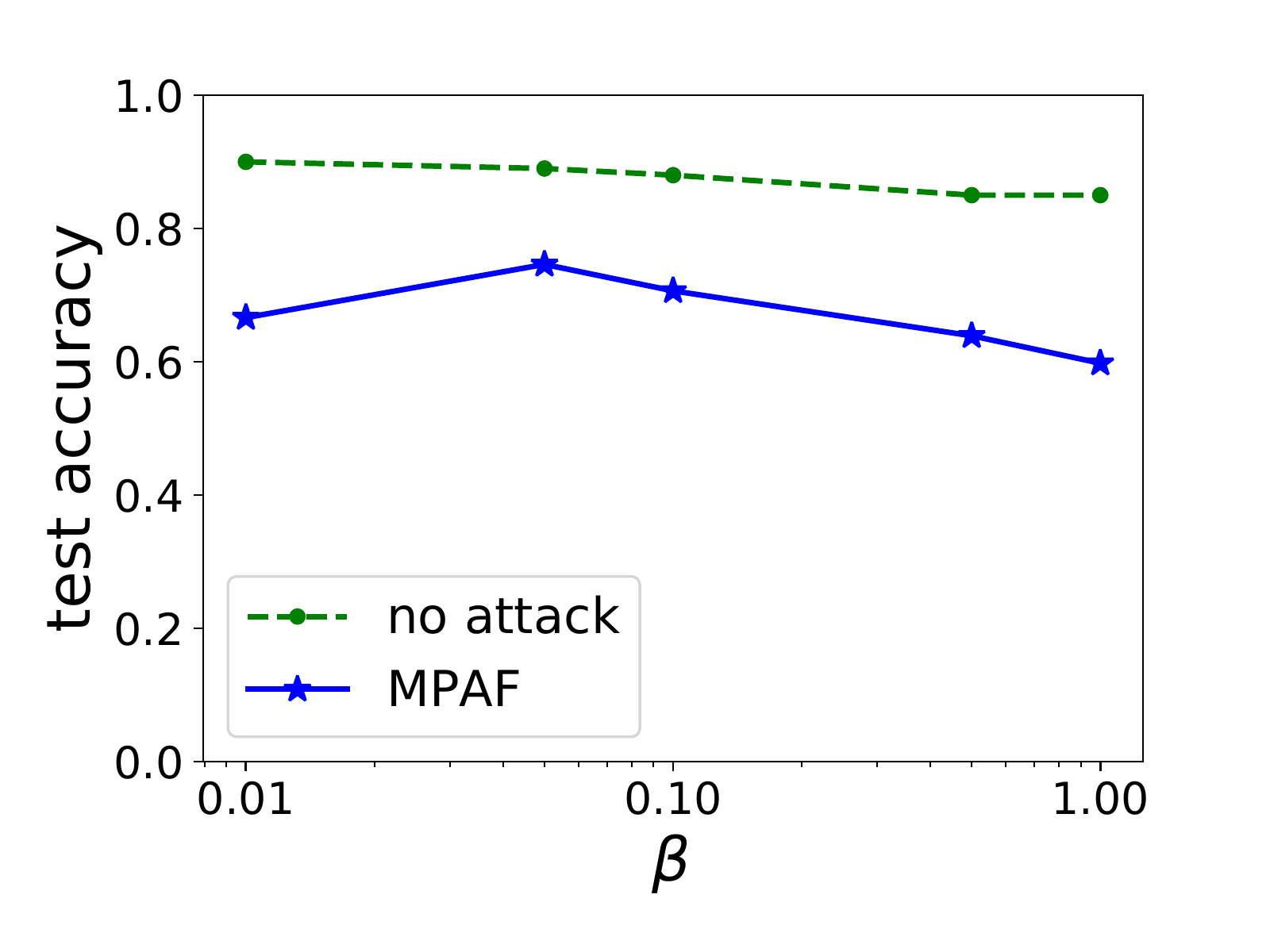}}
    \subfloat[Purchase]{\includegraphics[width=0.33\textwidth]{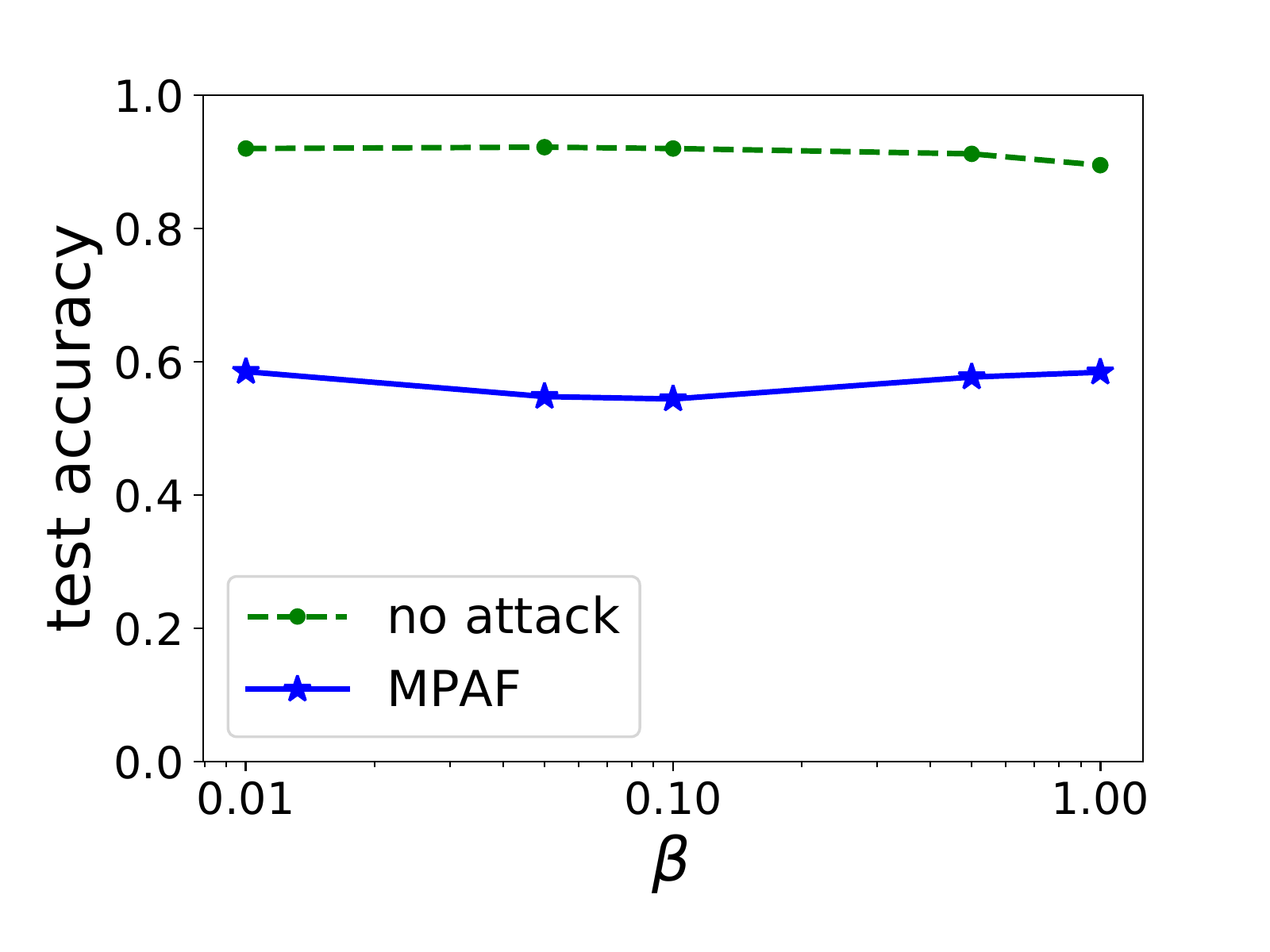}}
    \caption{Impact of the sample rate $\beta$ on the test accuracy of the global models learnt by Trimmed-mean. }
    \label{fig:beta}
\end{figure*}

\begin{figure*}[!ht]
    \center
    \subfloat[MNIST]{\includegraphics[width=0.33\textwidth]{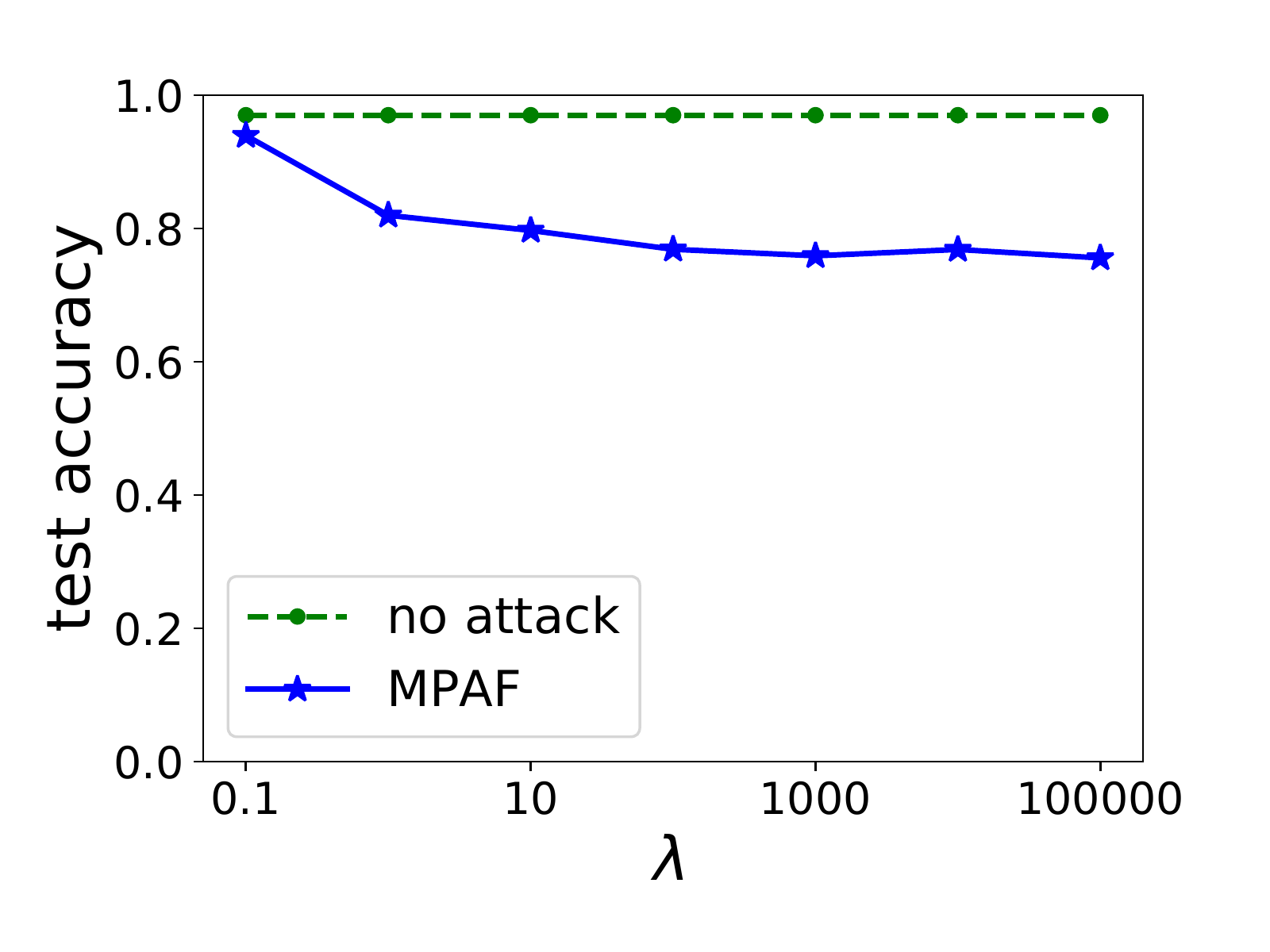}}
    \subfloat[Fashion-MNIST]{\includegraphics[width=0.33\textwidth]{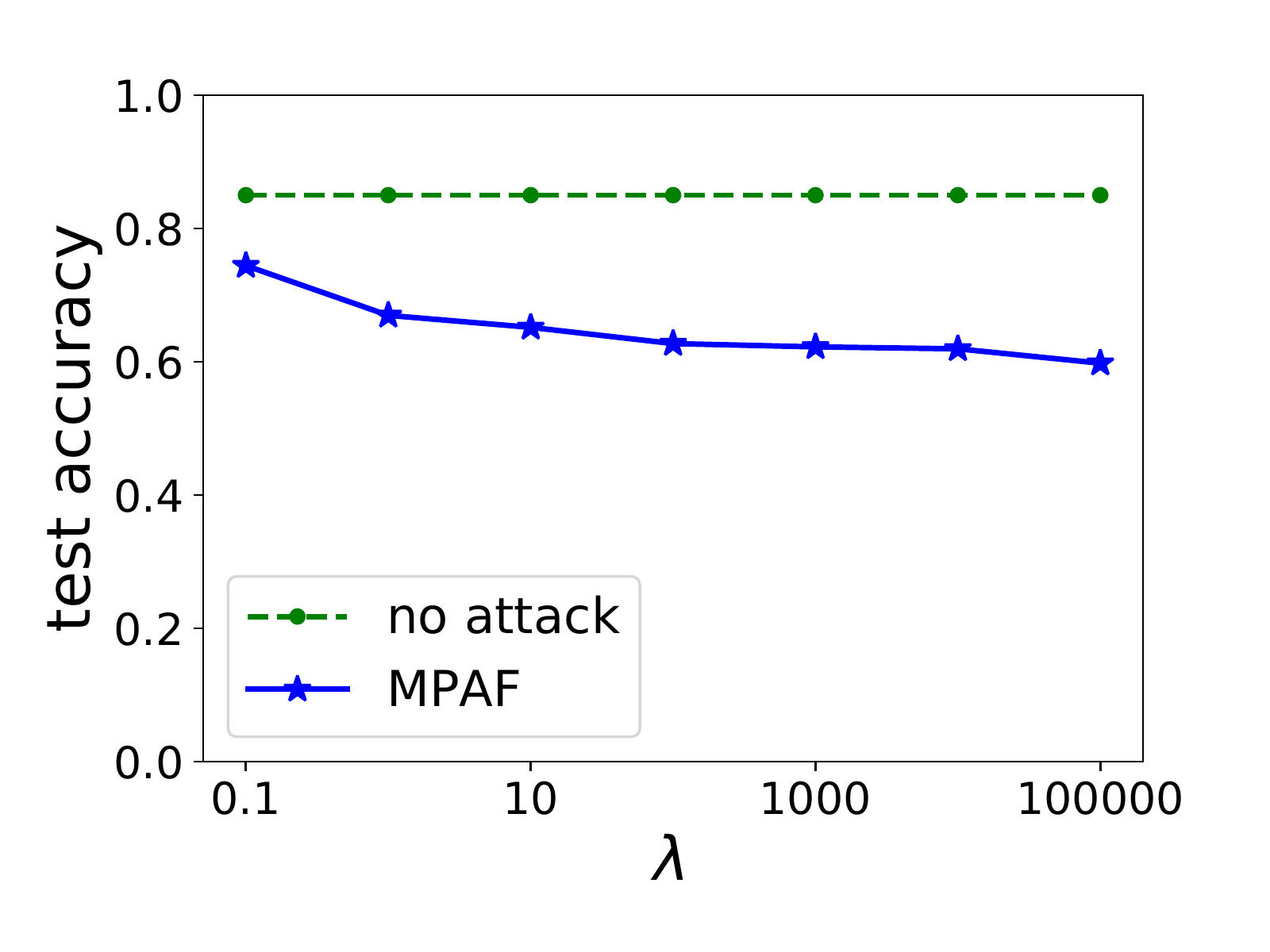}}
    \subfloat[Purchase]{\includegraphics[width=0.33\textwidth]{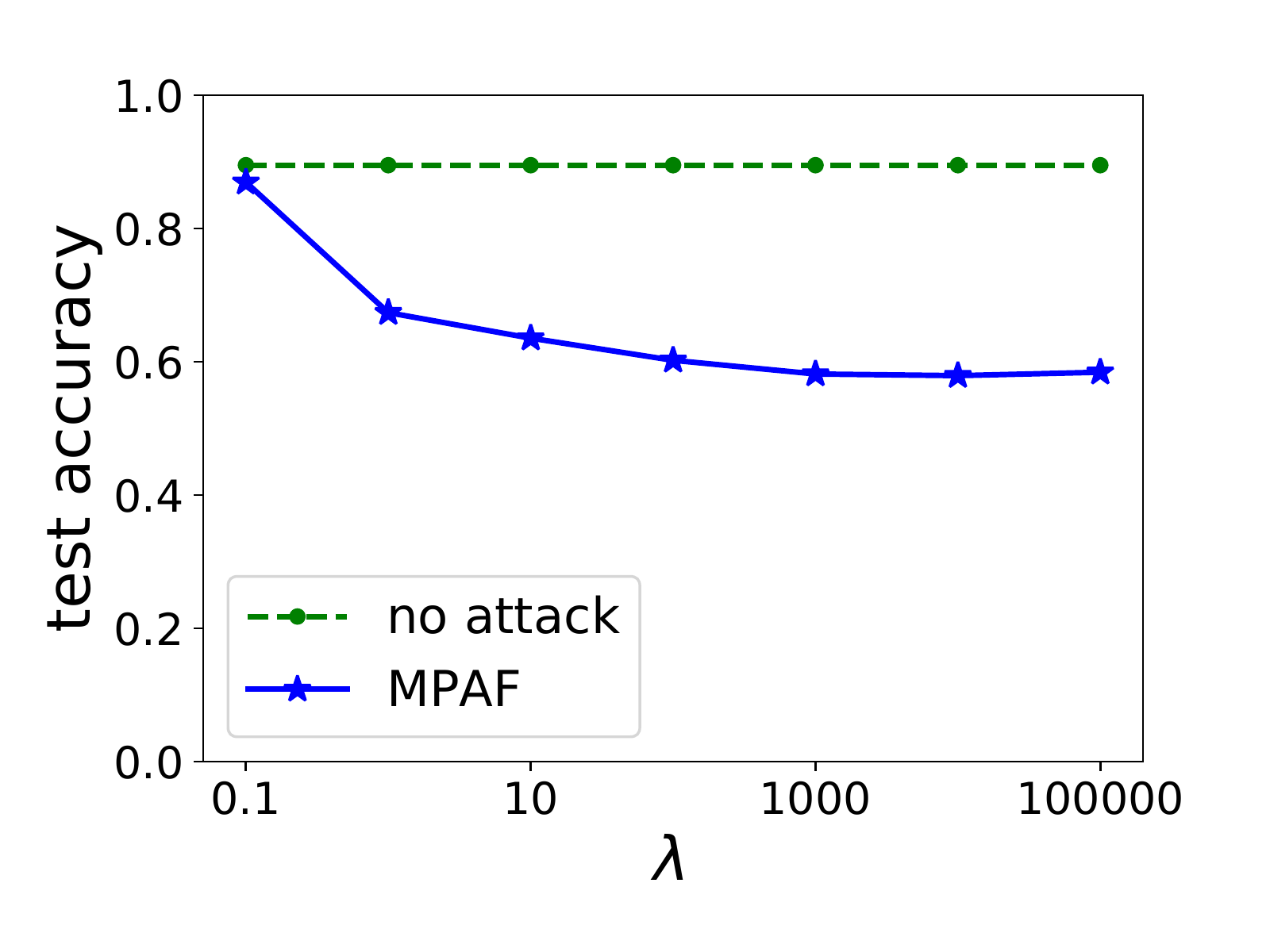}}
    \caption{Impact of the scaling factor $\lambda$ on the test accuracy of the global models learnt by Trimmed-mean.}
    \label{fig:lambda}
\end{figure*}

\subsection{Evaluation Results}
\myparatight{Impact of the fraction of fake clients} We explore the impact of the fraction of fake clients on two baseline attacks (i.e., random attack and history attack) and MPAF. Figure \ref{fig:fraction} shows the test accuracy of the global models learnt by different FL methods when the fraction of fake clients varies on the three datasets. We observe that when FedAvg is used, both baseline attacks and MPAF can reduce the test accuracy of the learnt global models to random guessing with only 1\% fake clients. However, when classical defenses (e.g., Median and Trimmed-mean) are applied, MPAF can still significantly decrease the test accuracy while the baseline attacks cannot. For instance, on Purchase dataset, MPAF reduces the test accuracy of the global model learnt with Trimmed-mean by 32\% when there are 10\% fake clients, while the baseline attacks can only decrease the test accuracy by at most 4\%. Moreover, we also observe that MPAF is more effective when the fraction of fake clients is larger. For instance, on Purchase dataset when Trimmed-mean is used, the test accuracy that MPAF can reduce increases from 32\% to 49\% when the fraction of malicious clients increases from 10\% to 25\%.

\myparatight{Impact of the sample rate $\beta$} We evaluate the effectiveness of MPAF when the server samples different fractions of clients in each FL round. Figure \ref{fig:beta} shows the test accuracy of the global models learnt with Trimmed-mean on all three datasets. We omit the results of non-robust FedAvg for simplicity as the test  accuracy is consistently close to random guessing under MPAF. We observe that the sample rate $\beta$ does not have much impact on MPAF and that MPAF can significantly decrease the test accuracy when $\beta$ ranges from 0.01 to 1.00. The previous claim that FedAvg and classical defenses are robust to untargeted model poisoning attacks when $\beta$ is small \cite{shejwalkarback} does not apply to our attack. This is because their claim is based on the assumption that  an attacker can only compromise a small fraction of genuine clients. 

\myparatight{Impact of the scaling factor $\lambda$}
We explore the impact of the scaling factor on MPAF. Figure \ref{fig:lambda} shows the test accuracy of the global models learnt by Trimmed-mean on all three datasets. We observe that the test accuracy first decreases as $\lambda$ increases, and then remains almost unchanged when $\lambda$ further increases. Our results show that even though the attacker does not know the hyperparameters of FL (e.g., the global learning rate $\eta$), by choosing a reasonably large value for $\lambda$, e.g., $\lambda\ge 1$ in our experiments, MPAF can reduce the test accuracy of the global model significantly. 
\section{Norm Clipping as A Countermeasure}
A recent work \cite{sun2019can} has proposed norm clipping as a countermeasure against backdoor attacks in federated learning. Specifically, the server selects a norm threshold $M$, and clips all local model updates whose $\ell_2$-norm is larger than $M$ such that their $\ell_2$-norm becomes $M$. The local model updates whose $\ell_2$-norms are no larger than $M$ remain unchanged. Formally, a local model update $\bm{g}$ becomes $\frac{\bm{g}}{\max(1, \Vert\bm{g}\Vert_2/M)}$ after norm clipping. The largest $\ell_2$-norm of the clipped local model updates is $M$. Therefore, the impact of the malicious local model updates will be limited. As a result, the backdoor attacks \cite{bagdasaryan2020backdoor} that rely on scaled local model updates will have lower attack success rate when norm clipping is adopted as a countermeasure.

We note that the idea of using norm clipping as a countermeasure is not limited to backdoor attacks. In fact, it may also be leveraged as a countermeasure against untargeted attacks that involve scaling. In MPAF, we use a scaling factor $\lambda$ to increase the impact of fake local model updates during aggregation. Therefore, it is intuitive to apply norm clipping as a countermeasure against MPAF. We empirically evaluate the effectiveness of MPAF when norm clipping is used as a countermeasure. Specifically, we use our default setting for Fashion-MNIST dataset and Trimmed-mean as the aggregation rule. Before using Trimmed-mean to aggregate the local model updates, we clip them with norm threshold $M$, where we vary the value of $M$ in our experiments. We omit the results of FedAvg for simplicity as the test  accuracy is consistently close to random guessing under MPAF.

\begin{figure}
    \centering
    \includegraphics[width=0.45\textwidth]{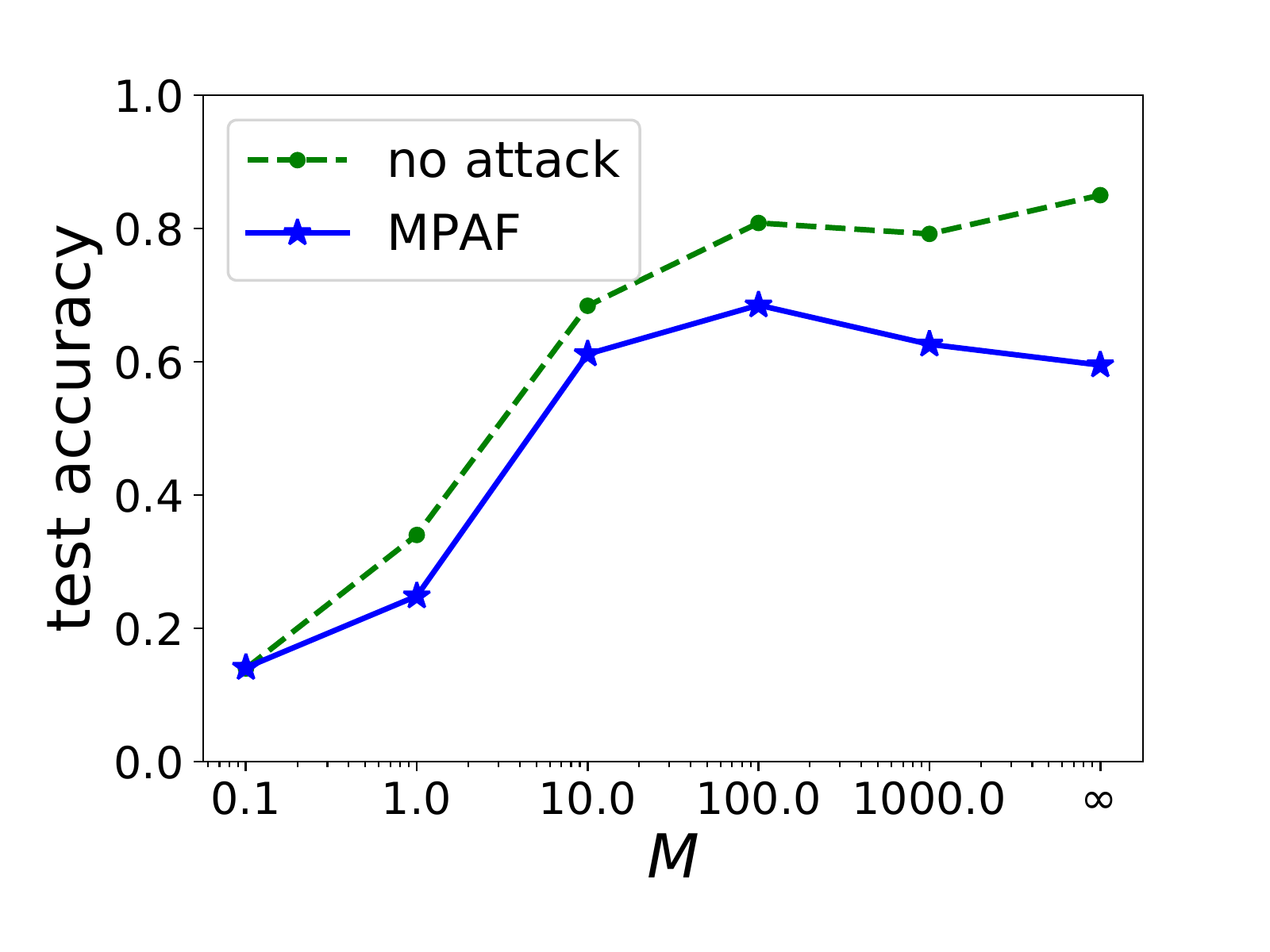}
    \caption{Impact of the norm clipping bound $M$ on the test accuracy of the global model learnt by Trimmed-mean on Fashion-MNIST.}
    \label{fig:fashion_M}
\end{figure}

Figure \ref{fig:fashion_M} shows the test accuracy of the global model learnt by Trimmed-mean on Fashion-MNIST. We use $M\rightarrow\infty$ to represent the case when there is no norm clipping. We observe that MPAF can still effectively decrease the test accuracy of the global model when norm clipping is deployed. Specifically, under no attack, the global model achieves the largest test accuracy of 0.85 when $M\rightarrow\infty$. However, under MPAF, the global model achieves the largest test accuracy of 0.68 when $M$ is around 100, which represents 0.17 accuracy loss. 
We also observe that the difference between the test accuracy of the global model under MPAF and the one under no attack is smaller as $M$ decreases. This is because more fake local model updates are clipped as $M$ decreases.  However, as $M$ decreases, the test accuracy under no attack also decreases, 
e.g., $M<100$ in Figure \ref{fig:fashion_M} leads to a test accuracy that is much lower than that when $M\rightarrow\infty$. This is because when $M$ decreases, more benign local model updates are also clipped, which results in a less accurate global model. Our results indicate that MPAF is still effective in reducing the test accuracy of the global model, even if both classical defenses (e.g., Trimmed-mean) and norm clipping are adopted.  
\section{Conclusion and Discussion}
In this work, we proposed MPAF, the first model poisoning attack to FL that is based on fake clients. We considered a minimum-knowledge setting for the attacker and showed that our attack is effective even when classical defenses and norm clipping are applied, highlighting the need for more advanced defenses against model poisoning attacks based on fake clients.

We hope our work can inspire more future studies on model poisoning attacks and their defenses. First, since it is unrealistic for an attacker to compromise a large fraction of genuine clients, it is more interesting to explore attacks based on fake clients. For instance, an interesting future work is to improve MPAF with extra knowledge, e.g., training data/model obtained from a similar learning task. 

Second, existing untargeted model poisoning attacks based on compromised genuine clients (e.g., \cite{fang2019local}) formulate round-wise optimization problems. Specifically, in each individual round of FL, the compromised genuine clients solve an independent problem to obtain the malicious local model updates. The solutions to these independent problems may contradict to each other. As a result, such malicious local model updates in different rounds may cancel each other out, leading to sub-optimal overall attack effect. On the contrary, our MPAF leverages a simple yet effective way of formulating a global optimization problem that deviates the global model towards a fixed base model. 

Third, it is an interesting future work to extend our MPAF to perform targeted model poisoning attacks. Specifically, an attacker can choose a base model that has an attacker-desired targeted behavior, e.g., a backdoored base model. 
By forcing the  learnt global model to be close to a backdoored base model, the learnt global model may have the same backdoor behavior as the base model and  predict attacker-chosen target labels for attacker-chosen test inputs.

\section*{Acknowledgements}
We thank the anonymous reviewers for constructive reviews and comments. This work was supported by National Science Foundation under grant No. 2112562 and 1937786, as well as Army Research Office under grant No. W911NF2110182.

\balance
{\small
\bibliographystyle{ieee_fullname}
\bibliography{refs}
}

\end{document}